\documentclass{aa}  
\usepackage{graphicx}
\usepackage{txfonts}
\usepackage{hyperref}
\usepackage{subfigure}
\usepackage{afterpage}
\usepackage{comment}
\usepackage{layouts}
\usepackage{multirow}
\usepackage{xcolor}

\newcommand*\diff{\mathop{}\!\mathrm{d}}

\newcommand{\zerodel}{.\kern-\nulldelimiterspace}

\newcommand{\SD}[1]{\textcolor{purple}{#1}}

\newcommand{\XMM}{XMM-\textit{Newton}~}

\begin{document} 

   \title{CHEX-MATE : turbulence in the ICM from X-ray surface brightness fluctuations} 

   \author{
        S. Dupourqué \inst{\ref{inst1}}
        \and 
        N. Clerc \inst{\ref{inst1}}
        \and
        E. Pointecouteau \inst{\ref{inst1}}
        \and 
        D. Eckert \inst{\ref{inst2}}
        \and
        M. Gaspari \inst{\ref{inst3}}
        \and
        L. Lovisari \inst{\ref{inst5}, \ref{inst12}}
        \and
        G.W. Pratt \inst{\ref{inst6}}
        \and
        E. Rasia \inst{\ref{inst7}, \ref{inst8}}
        \and
        M. Rossetti \inst{\ref{inst5}}
        \and
        F. Vazza \inst{\ref{inst9}, \ref{inst10}, \ref{inst11}}
        \and 
        M. Balboni \inst{\ref{inst5}, \label{inst13}} 
        \and 
        I. Bartalucci \inst{\ref{inst5}}
        \and
        H. Bourdin \inst{\ref{inst15}, \ref{inst16}}
        \and 
        F. De Luca \inst{\ref{inst15}, \ref{inst16}}
        \and
        M. De Petris \inst{\ref{inst14}}
        \and 
        S. Ettori \inst{\ref{inst11}}
        \and
        S. Ghizzardi \inst{\ref{inst5}}
        \and
        P. Mazzotta \inst{\ref{inst15}, \ref{inst16}}
          }

   \institute{IRAP, Université de Toulouse, CNRS, CNES, UT3-PS, Toulouse, France \\ \email{sdupourque@irap.omp.eu}\label{inst1} 
   \and Department of Astronomy, University of Geneva, Ch. d’Ecogia 16, CH-1290 Versoix, Switzerland\label{inst2} 
   \and Department of Astrophysical Sciences, Princeton University, 4 Ivy Lane, Princeton, NJ 08544-1001, USA\label{inst3} 
   \and Dipartimento di Fisica e Astronomia, Universit\'{a} di Bologna, Via Gobetti 93/2, 40122, Bologna, Italy\label{inst4} 
   \and Center for Astrophysics $|$ Harvard $\&$ Smithsonian, 60 Garden Street, Cambridge, MA 02138, USA \label{inst12} 
   \and Université Paris-Saclay, Université Paris Cité, CEA, CNRS, AIM, 91191, Gif-sur-Yvette, France \label{inst6} 
   \and INAF, Osservatorio di Trieste, v. Tiepolo 11, 34131, Trieste, Italy\label{inst7} 
   \and IFPU, Institute for Fundamental Physics of the Universe, Via Beirut 2, 34014 Trieste, Italy \label{inst8} 
   \and Istituto Nazionale di Astrofisica, IASF-Milano, via A. Corti 12, 20133, Milano, Italy\label{inst5} 
   \and Hamburger Sternwarte, University of Hamburg, Gojenbergsweg 112, 21029 Hamburg, Germany\label{inst9}
   \and Istituto di Radioastronomia, INAF, Via Gobetti 101, 40122, Bologna, Italy\label{inst10}
   \and INAF, Osservatorio di Astrofisica e Scienza dello Spazio, via Piero Gobetti 93/3, 40129 Bologna, Italy\label{inst11}
   \and DiSAT, Universit\`a degli Studi dell’Insubria, via Valleggio 11, I-22100 Como, Italy\label{inst13}
   \and Dipartimento di Fisica, Sapienza Università di Roma, Piazzale Aldo Moro 5, I-00185 Roma, Italy\label{inst14}
   \and Università degli studi di Roma ‘Tor Vergata’, Via della ricerca scientifica, 1, 00133 Roma, Italy\label{inst15}%
   \and INFN, Sezione di Roma 2, Università degli studi di Roma Tor Vergata, Via della Ricerca Scientifica, 1, Roma, Italy\label{inst16}% INAF - IASF Milano, via A. Corti 12, 20133 Milano, Italy 
}

   \date{Received ???; accepted ???}

% \abstract{}{}{}{}{} 
% 5 {} token are mandatory
 
  \abstract{The intra-cluster medium is prone to turbulent motion that will contribute to the non-thermal heating of the gas, complicating the  use of galaxy clusters as cosmological probes. Indirect approaches can estimate the intensity and structure of turbulent motions by studying the associated fluctuations in gas density and X-ray surface brightness. In this work, we want to constrain the gas density fluctuations at work in the CHEX-MATE sample to obtain a detailed view of their properties in a large population of clusters. To do so, we use a simulation-based approach to constrain the parameters of the power spectrum of density fluctuations, assuming a Kolmogorov-like spectrum and including the stochastic nature of the fluctuation-related observables in the error budget. Using a machine learning approach, we learn an approximate likelihood for each cluster. This method requires clusters to be not too disturbed, as fluctuations can originate from dynamic processes such as merging. Accordingly, we remove the less relaxed clusters (centroid shift $w>0.02)$ from our sample, resulting in a sample of 64 clusters. We define different subsets of CHEX-MATE to determine properties of density fluctuations as a function of dynamical state, mass and redshift, and investigate the correlation with the presence or not of a radio halo. We found a positive correlation between the dynamical state and density fluctuation variance, a non-trivial behavior with mass and no specific trend with redshift or the presence/absence of a radio halo. The injection scale is mostly constrained by the core region. The slope in the inertial range is consistent with Kolmogorov theory. When interpreted as originating from turbulent motion, the density fluctuations in $R_{500}$ yield an average Mach number of $\mathcal{M}_{3D} \simeq 0.4 \pm 0.2$, an associated non-thermal pressure support of $ P_{\text{turb}}/P_{\text{tot}} \simeq (9\pm 6) \%$ or a hydrostatic mass bias $b_{\text{turb}} \simeq 0.09 \pm 0.06$, which is in line with what is expected from the literature.}

   \keywords{X-rays: galaxies: clusters, galaxies: clusters: intracluster medium, turbulence}

   \maketitle
%
%-------------------------------------------------------------------

\section{Introduction}

Clusters of galaxies, which are the largest gravitationally bound structures in the Universe, house the majority of their baryonic content within the intracluster medium (ICM). This hot gas ($\sim 10^7-10^8$ K) quickly develops turbulent cascades as a result of the various perturbations it undergoes. Different phenomena will dictate the injection of kinetic energy into the ICM, depending on the location within the cluster. The central parts are dominated by feedback from the Active Galactic Nucleus \citep[AGN, ][]{mcnamara_mechanical_2012, gaspari_can_2014, voit_global_2017} while the outer parts are prone to accretion from the cosmic web and merging with groups or clusters of galaxies which will induce shocks and therefore turbulence \citep{nelson_evolution_2012}. These perturbations and their associated turbulence significantly contribute to the non-gravitational heating of the ICM, thereby influencing the hydrostatic mass bias of galaxy clusters \citep{piffaretti_total_2008, lau_residual_2009, rasia_lensing_2012, nelson_weighing_2014, biffi_nature_2016, pratt_galaxy_2019, gianfagna_study_2023}.

Generally speaking, turbulence is characterized by chaotic, irregular motion in a fluid, resulting from the injection of high-scale kinetic energy and high Reynolds numbers. It involves the transfer of energy from large scales to smaller scales within the fluid, leading to complex, unpredictable patterns of motion. { Turbulence in galaxy clusters is extensively modelled through numerical simulations \citep[e.g.][]{vazza_turbulence_2012, schmidt_hot_2016, zuhone_what_2018, mohapatra_turbulence_2019, ayromlou_atlas_2023}.} Studying turbulence in galaxy clusters can help in characterising the non-thermal pressure support of the ICM, as this should be one of its key components \citep{vazza_turbulent_2018, angelinelli_turbulent_2020}, together with magnetic fields and, to a lesser extent, cosmic rays. 

Constraining this process is relatively straightforward with highly resolved X-ray information in the spectral domain, as the motion of the gas is mirrored in the observed spectrum through centroid shifts and broadening of the emission lines. { Direct measurements of the line centroid shift were performed with the XIS instrument onboard \textit{Suzaku}, deriving upper limits as a subsonic line of sight motions in local clusters of various dynamical states \citep{ota_suzaku_2007, sugawara_suzaku_2009, tamura_discovery_2011, tamura_gas_2014, ota_search_2016}. More recently, by calibrating the EPIC-pn instrument of \textit{XMM-Newton} using the instrumental fluorescence lines, measurements of velocity maps with a precision of $\sim 200 - 500 $ km s$^{-1}$ were performed for local clusters, and the first centroid shift structure function from direct observations was derived \citep{gatuzz_measuring_2022, gatuzz_velocity_2022,gatuzz_measuring_2023-1, gatuzz_measuring_2023}. Broadening measurements were performed using the RGS instrument of \textit{XMM-Newton} \citep{sanders_constraints_2011, sanders_velocity_2013, pinto_chemical_2015, ogorzalek_improved_2017}, providing overall upper limits on subsonic motions. A comprehensive review of the study of the ICM motion is proposed by \citet{simionescu_constraining_2019}. Calibration uncertainties and limited spatial resolution are the main showstoppers when studying the ICM motions in the X-ray domain. The emergence of X-ray integral field units will enable high-precision direct measurements of such turbulent motions \citep[e.g.][]{roncarelli_measuring_2018, zhang_mapping_2023}. The \cite{hitomi_collaboration_quiescent_2016} determined a line-of-sight velocity dispersion of $\sim 160 \text{ km s}^{-1}$ in the core of Perseus cluster. Similar measurements will be accessible with the recent launch of \textit{XRISM}/Resolve \citep{xrism_science_team_science_2020}, the Line Emission Mapper \citep[LEM,][]{kraft_line_2022} and \textit{Athena}/X-IFU in the long term \citep{barret_athena_2020}}. Indirect measurements can be achieved by searching for fluctuations in the thermodynamical properties of the ICM that can be due to turbulent processes. For instance, X-ray surface brightness (X-SB) fluctuations point to a pure hydrodynamical flow in the ICM \citep{schuecker_probing_2004}. Such methods have been applied to the Coma \citep{churazov_x-ray_2012,gaspari_constraining_2013} and Perseus cluster \citep{zhuravleva_gas_2015}. \cite{zhuravleva_gas_2018} used a small sample of clusters to investigate the turbulent motions occuring in their cool-cores. \cite{hofmann_thermodynamic_2016} investigated the fluctuations of several thermodynamical properties of the ICM. \cite{khatri_thermal_2016} and \cite{romero_inferences_2023} showed a promising approach, joining the X-ray images to the Sunyaev-Zel'dovich effect (SZE). 

However, these studies do not account for the stochastic nature of the surface brightness fluctuations. Indeed, when observing a random field with a finite size sample and trying to assess its properties, an additional source of irreducible variance is added to the total error budget. As shown in \cite{clerc_towards_2019} and in \cite{cucchetti_towards_2019}, the so-called sample variance may be dominant at any spatial scale achievable by \XMM and its prevalence grows with scale up to an order of magnitude with respect to other sources. In \cite{dupourque_investigating_2023}, we performed an X-ray surface brightness fluctuation analysis on the X-COP cluster sample \citep{eckert_xmm_2017}, with a forward model approach to explicitly account for the sample variance. In this paper, we apply this approach to the CHEX-MATE cluster sample, which is an order of magnitude higher in terms of statistics, and contains a more diverse population of clusters.

In Sect.~\ref{sec:data_method}, we describe the methodology to compute the density fluctuation parameters. In Sect.~\ref{sec:results}, we present the joint constraints on density fluctuations, applied to various subsamples of CHEX-MATE, show the effect of excluding the central part of clusters in our analysis, and investigate the correlation with radio data using the LOw Frequency ARray (LOFAR) Two Meter Sky Survey second data release \citep[LoTSS DR2,][]{shimwell_lofar_2022,botteon_planck_2022}. In Sect.~\ref{sec:discussions}, we discuss the Mach number obtained when assuming that the density fluctuations originate from turbulent processes, and the resulting turbulent mass bias. Throughout this paper, we assume a flat $\Lambda$CDM cosmology with $H_0 = 70 \text{ km s}^{-1}$ and $\Omega_m = 1 - \Omega_\Lambda = 0.3$. Scale radii are defined according to the critical density of the Universe at the corresponding redshift. The $R_{500}$ and $M_{500}$ values are obtained from the MMF3 detection method \citep{melin_catalog_2006} as highlighted in Appendix A of \cite{the_chex-mate_collaboration_cluster_2021}. The Fourier transform conventions are highlighted in Appendix \ref{app:fourier_convention}.

\begin{comment}
\SD{
\begin{itemize}
    \item textwidth in inch: \printinunitsof{in}\prntlen{\textwidth}
    \item linewidth in inch: \printinunitsof{in}\prntlen{\linewidth}
\end{itemize}
}
\end{comment}

\section{Data and Method}
\label{sec:data_method}

\subsection{The CHEX-MATE Sample}

The CHEX-MATE program\footnote{\url{http://xmm-heritage.oas.inaf.it/}}, detailed in \cite{the_chex-mate_collaboration_cluster_2021} paper, is a three mega-second, multi-year \XMM Heritage Programme aimed at obtaining X-ray observations of 118 minimally biased, signal-to-noise-limited galaxy clusters detected by Planck through SZE \citep{the_planck_collaboration_planck_2016-1}. The project's objectives are to accurately understand the statistical properties of the cluster population, examine how gas properties are influenced by dark matter halo collapse, reveal non-gravitational heating origins, and address major uncertainties in mass determination that restrict the use of clusters for cosmological parameter estimation.

To achieve these goals, a sample of 118 Planck clusters was chosen based on their SZE signal (S/N > 6.5) and categorized into two subsamples: Tier 1 contains 61 low-redshift objects in the northern sky that provide an unbiased view of the most recent cluster population ($0.05 < z < 0.2$ and $ 2 < \frac{M_{500}}{10^{14} M_{\odot}}$ < 9); Tier 2 is representative of  the most massive systems formed in the Universe's history ($z < 0.6$ and $\frac{M_{500}}{10^{14} M_{\odot}} > 7.25 $). These subsamples share four common clusters. The \XMM observations are characterized by an exposure time ensuring an S/N of 150 within $R_{500}$ in the [0.3-2.0] keV band. This requirement was chosen to satisfy three conditions: i) to calculate the temperature profile up to $R_{500}$ (with an accuracy of $\pm 15\%$ in the [0.8–1.2] $R_{500}$ region), ii) to measure the mass obtained from the $Y_X$ mass proxy \citep{kravtsov_new_2006}, in which $Y_X$ = $M_{g,500} T_X$ ($M_{g,500}$ represents the gas mass within $R_{500}$ and $T_X$ refers to the spectroscopic temperature estimated in the [0.15-0.75] $R_{500}$  range), with an uncertainty of $\pm 2 \%$, and iii) to derive the mass derived from hydrostatic equilibrium (HE) at $R_{500}$ with a precision of approximately $15-20\%$.

For further information on the sample, scientific objectives, and observation strategies in X-ray and other wavebands, readers should consult the presentation paper: \cite{the_chex-mate_collaboration_cluster_2021}.

\subsection{Data preparation}
\label{sec:data_preparation}

The images in this study were generated using the pipeline established for the X-COP sample \citep[][]{eckert_xmm_2017, ghirardini_universal_2019} and adopted by the CHEX-MATE collaboration. Specifically, \XMM data were processed with SAS software (version 16.1.0) and the Extended Source Analysis Software package \citep[ESAS,][]{snowden_catalog_2008}. { The choice of SAS version 16.1 is motivated in Rossetti et al. 2023 (submitted) to which we refer for more details. In brief, we aim to avoid the presence of unresolved bugs in intermediate versions of SAS and a major refactoring effort for version 21.0, which makes existing reduction pipelines incompatible for the time being. However, we ensured that the latest calibration available in January 2021 was applied, as specified by \cite{bartalucci_chex-mate_2023}.} Count images, exposure maps, and particle background maps were extracted in the narrow [0.7-1.2] keV band to maximize the source-to-background emission ratio and minimize systematics related to EPIC background subtraction. A detailed account of the procedure is presented in \cite{bartalucci_chex-mate_2023}, and a complete gallery of images is displayed in Fig. 6 in \cite{the_chex-mate_collaboration_cluster_2021}. Point sources are extracted using the \texttt{ewavelet} routine in the two energy bands [0.3-2] and [2-7] keV. Every source with flux smaller than the maximum of the logN-logS distribution determined from the point source extraction is masked.

\subsection{Bayesian inference for density fluctuations}

The methodology described below was first developed and discussed in detail in \cite{dupourque_investigating_2023}. This is a 3-step procedure that consists of determining an unperturbed surface brightness model, defining the associated surface brightness fluctuations and deriving a meaningful observable, and finally constraining parameters of the 3D power spectrum, including the whole error budget. We summarise it briefly in the following sections but refer interested readers to the aforementioned publication for further details.

\subsubsection{X-SB unperturbed model}

In this study, we aim to model the unperturbed surface brightness $S_X$ of a galaxy cluster to define fluctuations. Using $\vec{r} = (x, y, \ell)$ as the 3D position parametrization, and $\vec{\rho} = (x,y)$ as the 2D equivalent, the surface brightness can be expressed as:

\begin{equation*}
S_{X}(\Vec{\rho}) = \int^{+\infty}_{-\infty}\Psi(\vec{r}) \: n_e^2(\vec{r}) \: \diff \ell + B,
    \label{eq:surface_brightness_definition}
\end{equation*}

where $n_e(\Vec{r})$ is the electronic density and $\Psi(\vec{r})$ represents a combination of various factors such as the cooling function, cosmological dimming, Galactic absorption, and convolution with the XMM response functions. $B$ denotes a constant surface brightness sky background. This radial model is folded in a triaxial shape for the cluster, with no scaling along the line of sight. The centre position and ellipticity are left free, and the centre is fitted with a normal prior centred on the X-ray peak as defined by \cite{bartalucci_chex-mate_2023}, with a 50\% relative scatter. A modified Vikhlinin model \citep[with $\alpha = 0$ and $\gamma = 3$, see][]{vikhlinin_chandra_2006, shi_analytical_2016} is used to parametrize the density profile:

\begin{equation*}
    n_e(r) = n_{e,0} \frac{(1+r^2/r_c^2)^{-3\beta/2}}{(1+r^3/r_s^3)^{\epsilon / 6}}.
    \label{eq:vikhlinin_density_profile}
\end{equation*}

{ The $\Psi(\vec{r})$ function is determined using an analytical functional form that we fit on a count-rate grid estimated using \texttt{XSPEC 12.11.1}. To do so, we compute the expected count-rate of a \texttt{PhAbs*APEC} model with the same instrumental setup and redshift for each cluster in the CHEX-MATE sample, more details are available in App. C of \cite{dupourque_investigating_2023}. The metallicity was set to $0.3~Z_{\odot}$ and the abundances fixed according to \cite{anders_abundances_1989}.} Galactic absorption is accounted for using the $N_H$ data from the HI4PI survey \citep{hi4pi_collaboration_hi4pi_2016}. The count image is rebinned using Voronoi tessellation \citep{cappellari_adaptive_2003}, with approximately 100 counts per bin, to increase the speed of the inference given the computationally intense surface brightness model that we consider. This binning introduces a median bias of $\sim 0.5 \%$ which peaks at 4 $\%$ in the outermost (and low significance) region, which is still lower than the expected $10 \%$ systematic error due to Poisson noise, see App. D in \cite{dupourque_investigating_2023} for more details. The number of counts expected in each bin is forward modelled using the aforementioned surface brightness model. The model parameters are determined using Bayesian inference with a Poisson likelihood in each bin. The posterior distributions are sampled using the No U-Turn Sampler \citep{hoffman_no-u-turn_2014} as implemented in the \texttt{numpyro} library \citep{bingham_pyro_2019, phan_composable_2019}.

\subsubsection{Surface brightness \& density fluctuations }

Surface brightness emission from galaxy clusters in the soft X-ray band originates mainly from the thermal bremsstrahlung occurring in the ICM plasma. The intensity of this radiation is directly proportional to the squared electronic density. Hence, over-densities and under-densities in the gas distribution can be found as fluctuations in the surface brightness images, and dominate the fluctuations when compared to other sources \citep{churazov_x-ray_2012}. To constrain the properties of the density fluctuations, we choose to model it as a Gaussian random field, which can be described using only a second order moment such as the power spectrum or the structure function. As we expect a strong link between density fluctuations and turbulence \citep{gaspari_constraining_2013, gaspari_relation_2014, zhuravleva_relation_2014, simonte_exploring_2022}, we model the 3D power spectrum of the random field using a Kolmogorov-like functional form \citep[e.g.][]{zuhone_mapping_2016, dupourque_investigating_2023} since turbulence in clusters is compatible with the purely hydrodynamical case \citep{schuecker_probing_2004}:

\begin{equation}
    \mathcal{P}_{3D, \delta}(k)= \sigma_\delta^2 \frac{e^{-\left(k/k_{\text{inj}}\right)^2} e^{-\left(k_{\text{dis}}/k\right)^2} k^{-\alpha} }{\int 4\pi k^2  \, e^{-\left(k/k_{\text{inj}}\right)^2} e^{-\left(k_{\text{dis}}/k\right)^2} k^{-\alpha} \diff k},
\label{eq:p3dmodel}
\end{equation}

where $k_{\text{inj}}$ and $k_{\text{dis}}$ are, respectively, the Fourier frequencies corresponding to the injection scale $\ell_{\text{inj}}$ and dissipation scale $\ell_{\text{dis}}$, $\alpha$ is the inertial range spectral index (shortened to “slope”), and $\sigma_\delta^2$ is the variance of fluctuations. We fix the dissipation scale for our entire sample at $10^{-3} ~R_{500}$ \citep[e.g.][]{lazarian_turbulence_2015}, which is significantly smaller than the spatial scales accessible with CHEX-MATE ($\gtrsim 10^{-2} ~R_{500}$. In any case, the arbitrary choice of dissipation scale has little to no impact on the normalization of the power spectrum. 

\subsubsection{X-SB fluctuations and power spectrum}

In this analysis, the  X-SB fluctuations are assumed to originate exclusively from intrinsic density fluctuations \citep[see][for other sources of X-SB fluctuations]{churazov_x-ray_2012}. The squared density profile can be split into a rest component $n_{e,0}^2$ and fluctuations $\delta << 1$, such that the density is given by $n_e(\vec{r}) = n_{e, 0}(\vec{r}) \times (1+ \delta(\vec{r}))$. Using a first order expansion of the density, the raw surface brightness image $S_{X}(\Vec{\rho})$ can be decomposed as the sum of an unperturbed image $S_{X,0}(\Vec{\rho})$ and a surface brightness fluctuation map. We define the fluctuation map as follows: 

\begin{equation}
    \Delta(\Vec{\rho}) \overset{\operatorname{def}}{=} \frac{S_{X}(\Vec{\rho}) - S_{X,0}(\Vec{\rho})}{2} \simeq\int^{+\infty}_{-\infty} \epsilon_0(\Vec{r})\delta(\Vec{r}) \diff \ell
\label{eq:sb-fluc-diff}
,\end{equation}

where $S_{X,0}$ can be obtained using the best-fit X-SB model and $\epsilon_0(\Vec{r}) = \Psi(\Vec{r})~ n_{e, 0}^2(\Vec{r})$ is the associated best-fit emissivity. The factor two arises when considering the square dependency of $\delta$ with the surface brightness. To study the contribution of each spatial scale to surface brightness fluctuations, an analysis in Fourier space is performed. We note the Fourier 3D frequencies as $\Vec{k} = (k_x, k_y, k_\ell)$, and associated 2D frequencies as $\Vec{\kappa} = (k_x, k_y) = (\kappa, \varphi_k)$ in cylindrical coordinates. The power spectrum of the fluctuations, $\mathcal{P}_{2D,\Delta}$, is defined as:

\begin{equation*}
    \mathcal{P}_{2D, \Delta}(\kappa) = \frac{1}{2\pi} \int |\Hat{\Delta}(\Vec{\kappa}) |^2 \diff \varphi_k
,\end{equation*}

where $\Hat{\Delta}$ is the Fourier transform of the two-dimensional (2D) map $\Delta$ and $\varphi$ is the direction angle in Fourier space. The numerical evaluation of $P_{2D,\Delta}$ is performed using the method described by \cite{arevalo_mexican_2012}, which computes the variance of images filtered by Mexican hats on a characteristic scale to estimate the power spectrum, implicitly handling irregular masks such as the excluded point sources or \XMM mosaics. In practice, we divide the X-SB fluctuation power spectrum $\mathcal{P}_{\text{2D, }\Delta}$ by the power spectrum of the best-fit X-SB model with Poisson noise $\mathcal{N}_{\text{2D, }\Delta}$, to isolate exclusively the fluctuations due to density variations. This observable is defined as a Signal-to-Noise ratio (S/N) as follows: 

\begin{equation}
\label{eq:snr-definition}
    \text{S/N}_{\Delta} \overset{\operatorname{def}}{=} \frac{\mathcal{P}_{\text{2D, }\Delta}}{\mathcal{N}_{\text{2D, }\Delta}}.
\end{equation}

We compute this spectrum on scales defined by the size of the cluster in $R_{500}$ units. The observational strategy detailed in \cite{the_chex-mate_collaboration_cluster_2021} and Rossetti et al. 2023 (submitted) ensures that the thermodynamic properties can be reconstructed up to $R_{500}$, so we use photons inside this radius. Since the CHEX-MATE sample has a wide range of redshifts, the lowest scale we can investigate is greater for distant clusters than for nearby clusters. We choose a conservative, redshift-dependent scale that allows us to study all clusters respecting the Nyquist-Shannon criterion and minimizing the lost information, in the form of an affine function: 

\begin{equation}
\label{eq:low-scale}
    \rho_{\text{low}} \simeq  0.123\times z + 0.023 ~ \left[R_{500}\right].
\end{equation}

\begin{figure}
    \centering
    \includegraphics[width=\hsize]{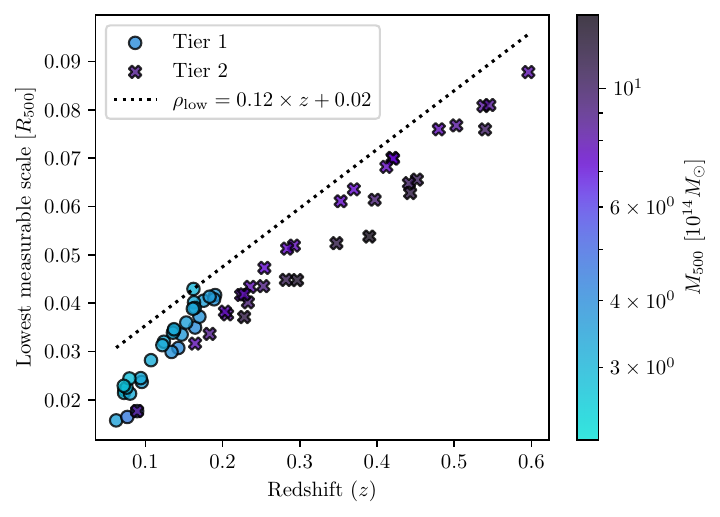}
    \caption{Lowest scale accessible as a function of the redshift, and best conservative affine relationship used in this paper as defined in Eq.~\ref{eq:low-scale}, which is defined as an affine function with the same slope as the points, but raised to be above them.}
    \label{fig:low_scale_redshift}
\end{figure}

The lowest scale for each cluster in the sample as a function of $z$ is displayed in Fig.~\ref{fig:low_scale_redshift}, where Eq.~\ref{eq:low-scale} is shown with a dotted line. The effect of this redshift-dependent scale will be to reduce the importance of distant clusters compared with closer ones, since the information linked to them is intrinsically less constraining.

%The power spectrum is finally evaluated in a region between $0.15$ and $1~R_{500}$ in order to mitigate the central morphological effects associated with the presence of cool-cores or sloshing.

\subsubsection{Simulation-based inference}

To link our observable quantities to the parameters of the density fluctuations, we need to establish the formal link between these two quantities, which is encapsulated by the likelihood function in a Bayesian framework. However, we cannot derive an analytical formula or closed form for this likelihood in this case. It is however possible to generate mock observables, by generating a Gaussian random field with known parameters and injecting it into our best-fit emissivity model before projection along the line of sight. By including all our sources of uncertainty in this forward modelling, it is possible to assess the dispersion associated with our parameters and our observable due to the sample variance and Poisson noise. { This procedure is repeated for 300,000 simulations per cluster to build the training sample, whose generating parameters are randomly drawn from the following priors: $\sigma_{\delta} \sim 10^{\mathcal{U}\left(-2, 0\right)}$, $\ell_{\text{inj}} \sim 10^{\mathcal{U}\left(-2, 0.3\right)}$ and $\alpha \sim \mathcal{U}\left(2, 5\right)$. We showed in \cite{dupourque_investigating_2023} that this number of 
simulations is sufficient to build a likelihood estimator enabling reconstruction of the expected parameters for mock observations}. We train a neural network on these simulations to estimate an approximation of the likelihood of our problem, using a masked autoregressive flow \citep{papamakarios_masked_2017, papamakarios_sequential_2019} as implemented in the \texttt{sbi} package \citep{tejero-cantero_sbi_2020}. { In a nutshell, a masked autoregressive flow is a density estimator that is adjusted to estimate any well-behaved distribution (in the mathematical sense). In this problem, the likelihood of a specific observation is learned from the 300,000 joint samples of parameters and mock observations. Once it is trained, the model can return an approximation of the likelihood of any observation. This approximate likelihood is sampled and inverted into a posterior distribution using classical MCMC methods such as NUTS \citep{hoffman_no-u-turn_2014}.}

\subsection{Splitting CHEX-MATE into subsamples}

\begin{table}
    \centering
    \begin{tabular}{c|c|c}
        Subsample & Definition & Size\\
        \hline
        \hline
State (I) & $0.001<w<0.005$ & 22\\
State (II) & $0.005<w<0.011$ & 21\\
State (III) & $0.011<w<0.019$ & 21\\
\hline
Mass (I) & $2.2<M_{500}<3.77$ & 11\\
Mass (II) & $3.77<M_{500}<4.65$ & 10\\
Mass (III) & $4.65<M_{500}<8.77$ & 10\\
\hline
Redshift (I) & $0.09<z<0.26$ & 12\\
Redshift (II) & $0.26<z<0.42$ & 11\\
Redshift (III) & $0.42<z<0.6$ & 11\\
\hline
Radio halo & Halo in LoTSS DR2 & 8\\
No radio halo & No halo in LoTSS DR2 & 13\\

    \end{tabular}
    \caption{Definition of the subsamples used in the analysis and relative size of each sample. The total number of clusters used in the analysis is 64. Masses are in $10^{14} M_{\odot}$ unit.}
    \label{tab:subsamples_definition}
\end{table}

The presented methodology allows the estimation of the likelihood of individual observations of each cluster. By joining them, one can estimate the parameters of the density fluctuations on selected subsets of the CHEX-MATE sample. For this study, we exclude some clusters for quality or modelling reasons. Our modelling is effective on regular clusters. Disturbed clusters, which have strong departures from spherical symmetry, produce very inhomogeneous density fluctuations due to structural residues. It is important to emphasize that the descriptive power of the Gaussian field model decreases in clusters featuring structures resulting from its dynamic assembly, such as sloshing spirals or cool cores. To mitigate this effect, we want to systematically exclude clusters for which this approach is not suited, and use morphological indicators to do so. \cite{campitiello_dynamical_2021} extracted morphological parameters of the CHEX-MATE sample and compared it to a visual classification in three classes: i) relaxed, ii) mixed, iii) disturbed. The centroid shift $w$, which relates to the variation of the distance of the peak of luminosity and the centroid of the emission for a varying aperture up to $R_{500}$, was computed for every cluster in the CHEX-MATE sample, and shows good separation of the three states. This quantity is dimensionless as it is scaled to $R_{500}$. We choose to exclude clusters such as $w>0.02$, which is comparable to the threshold proposed by \cite{lovisari_x-ray_2017} to classify disturbed clusters, and correspond to 43 objects in the whole CHEX-MATE sample. \cite{campitiello_dynamical_2021} also derived a combined $M$ parameter \citep[see also][]{rasia_x-ray_2013, deluca_three_2021}, whose purpose was to estimate the disturbance of clusters on a continuum using a combination of multiple morphological indicators. Using $w$ or $M$ yields similar results in this study, as both can define a threshold to exclude the most visually disturbed clusters. However, using $w$ instead of $M$ removes more mixed clusters from the sample with the same threshold definition, which is arguably a more conservative choice. Finally, $w$ correlates better with the variance of density fluctuations than $M$ in the whole sample, with a Pearson $R$ of respectively 0.5 and 0.36, showing that $w$ is, in this case, more suited to track the surface brightness disturbances. It is worth noting that despite this threshold, some disturbed clusters remain in our sample, such as PSZ2G266.04-21.25, the famous “Bullet Cluster”, using either criteria to perform the selection. Along with this disturbance threshold, we have systematically excluded double clusters, as well as clusters that are contaminated by the galactic emission \citep[in particular CIZA clusters, see][]{ebeling_ciza_2000}, Virgo emission, or by clusters in the foreground or background. 

\begin{figure}
    \centering
    \includegraphics[width=\hsize]{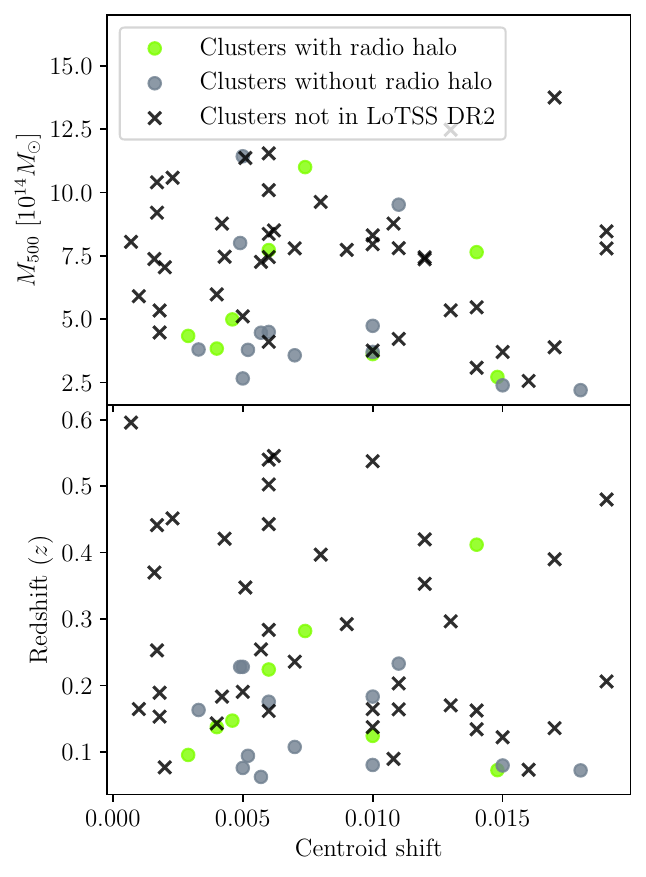}
    \caption{Correlation between the mass $M_{500}$, the redshift $z$ and the centroid shift $w$ in the CHEX-MATE sample without disturbed or contaminated clusters, including the clusters in LoTSS DR2 with and without radio halos.}
    \label{fig:mixing-between-sates}
\end{figure}

Applying this selection leads to a set of 64 clusters, which is $\sim 60\%$ of the total cluster number in CHEX-MATE. This set is divided into several subsamples, which are defined in Tab.~\ref{tab:subsamples_definition} and displayed in Fig.~\ref{fig:subsamples_definition}. The symbols (I), (II), (III) for each subsample denote an increase in the quantity used to define it, for instance, the State (I) contains the most relaxed clusters while the State (III) contains the most disturbed. In particular, we are interested in the dependence of the density fluctuation parameters on the mass of the clusters, their redshift, and their dynamical state. The sample is divided into bins of almost equal sizes to have comparable statistics. The CHEX-MATE sample is built on two subsamples, Tier 1 and Tier 2, which are respectively made up of close clusters of varying mass, and high mass clusters of varying redshifts. Hence, we perform mass separation using Tier 1 and redshift separation using Tier 2. The dynamical state separation is performed on the whole sample based on the parameter $w$. The CHEX-MATE sample is also partially observed by LOFAR. We use the radio halo data and power from the LoTSS-DR2 \citep{botteon_planck_2022}. This survey covers 40 clusters from the CHEX-MATE sample, in which 18 admit detected radio haloes. We apply the same filtering procedure and exclude the most disturbed clusters as done previously, lowering the number of clusters to 8 with a halo and 13 without a halo. To investigate the correlation between X-SB fluctuations and radio data, we further define two subsamples of clusters containing or not a radio halo. The good mixing between mass, redshift, dynamical state, and the presence of radio halo is shown in Fig.~\ref{fig:mixing-between-sates}, guaranteeing that we are not introducing any selection bias with our subsample definition.

\section{Results}
\label{sec:results}

\subsection{Joint constraints on the sample} \label{sec:constraints}

\begin{figure*}[t]
    \centering
    \includegraphics[width=0.45\hsize]{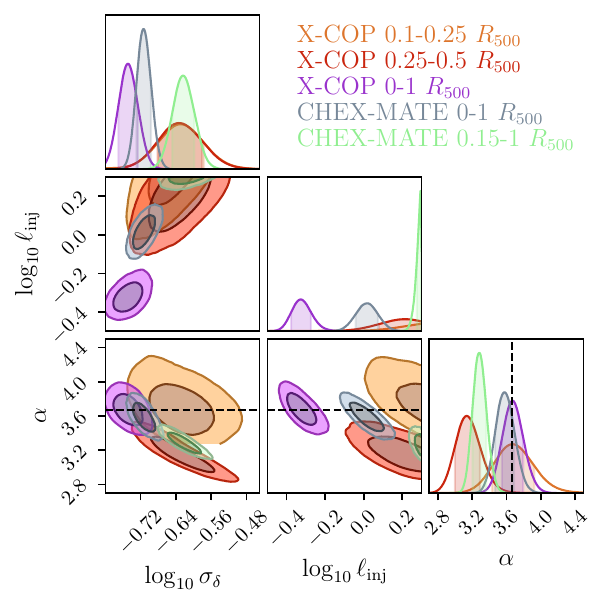}
    \includegraphics[height=0.45\hsize]{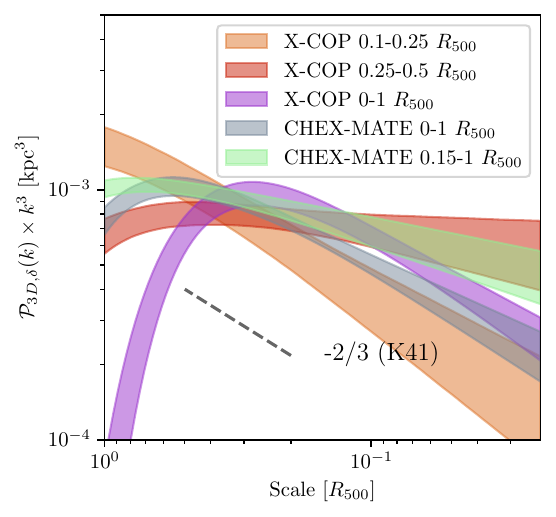}
    \caption{(\textit{left}) Joint posterior distributions of the standard deviation, $\sigma_\delta$, the injection scale, $\ell_\text{inj}$, and the spectral index $\alpha$ of the density fluctuation power spectrum parameters, estimated on the cleaned CHEX-MATE sample and with the X-COP parameters in the 0.1-0.25 and 0.25-0.5 $R_{500}$ rings, as computed in \cite{dupourque_investigating_2023}. The black dashed line represents the expected 11/3 index from Kolmogorov theory. (\textit{right}) Associated 3D power spectra as defined from Eq.~\ref{eq:p3dmodel}. The dashed slope indicates the expected slope from Kolmogorov. { The scales are in units of $R_{500}$.}}
    \label{fig:joint_with_xcop}
\end{figure*}

The constraints on surface brightness fluctuations can be significantly affected by sample variance when each cluster is studied individually. To mitigate this effect and increase the overall signal, we combine the analysis of density fluctuation properties across multiple clusters by summing their log-likelihoods. We calculate the X-SB fluctuation spectra using information within the $R_{500}$ radius. As cluster cores are often contaminated by cool-cores and structural residuals, we investigate the impact of excluding the central part up to 0.15 $R_{500}$ from our analysis. The joint parameters of the density fluctuations and their associated power spectra, estimated across the CHEX-MATE sample without disturbed clusters (64 objects), are shown in Fig.~\ref{fig:joint_with_xcop} and in Tab.~\ref{tab:mach-number}, Tab.~\ref{tab:mach-number-without-center}, both with and without core exclusion. These are compared with the previous results on the X-COP sample, including the constraints from \cite{dupourque_investigating_2023} for the two rings between 0.1-0.25 and 0.25-0.5 $R_{500}$ as well as the entire $R_{500}$ area. The results obtained for the fluctuation variance $\sigma_\delta$ and the slope $\alpha$ of CHEX-MATE are structurally compatible with the X-COP 0.1-0.25 and 0.25-0.5 $R_{500}$ rings, except for the injection scale $\ell_{\text{inj}}$. This is further discussed in Sec.~\ref{sec:core-exclusion}. The slope of the CHEX-MATE analysis including the core, as in X-COP, is fully compatible with \citet[][K41]{kolmogorov_local_1941} theory.

\begin{figure*}[!ht]
\centering
\subfigure[Subsamples in $M_{500}-z$ plane]{
\includegraphics[width=0.31\hsize]{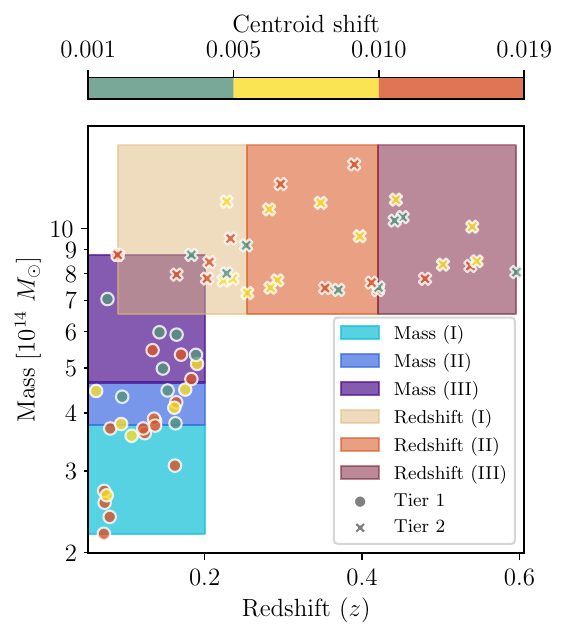}
\label{fig:subsamples_definition}
}
\subfigure[Full sample with core]{
\includegraphics[width=0.31\hsize]{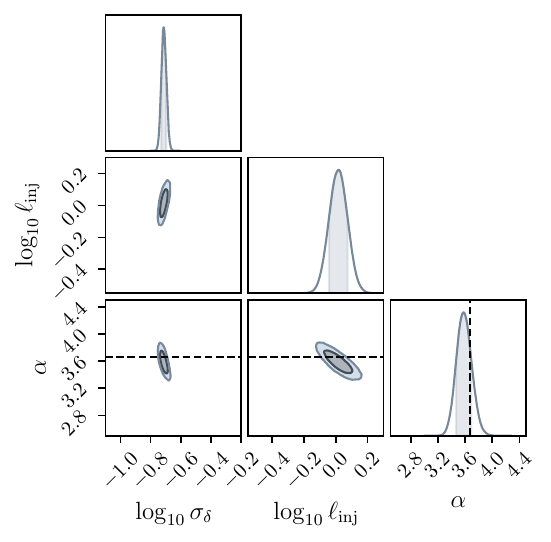}
\label{fig:with_core}
}
\subfigure[Full sample without core]{
\includegraphics[width=0.31\hsize]{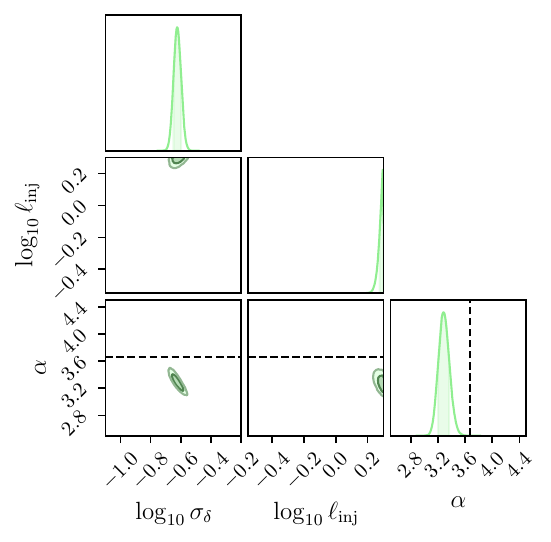}
\label{fig:without_core}
}
\subfigure[Splitting by states (with core)]{
\includegraphics[width=0.31\hsize]{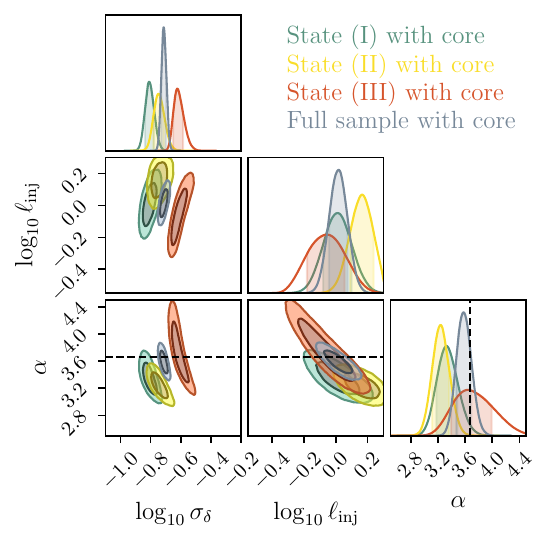}
\label{fig:joint_state}
}
\subfigure[Splitting by masses (with core)]{
\includegraphics[height=0.31\hsize]{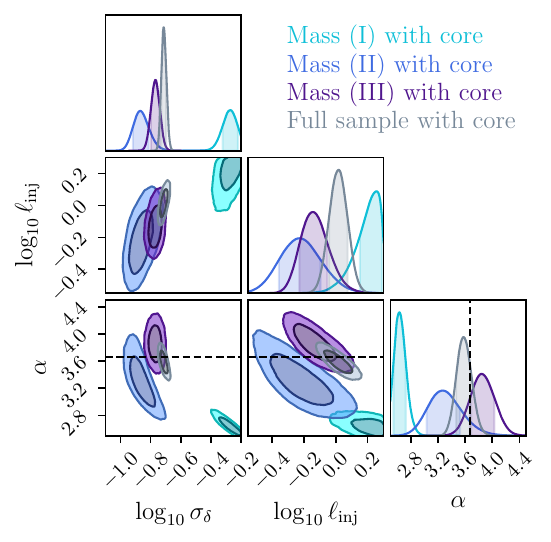}
\label{fig:joint_mass}
}
\subfigure[Splitting by redshifts (with core)]{
\includegraphics[width=0.31\hsize]{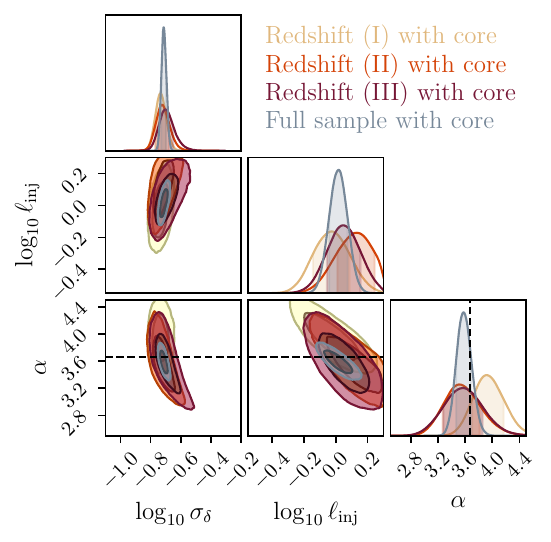}
\label{fig:joint_z}
}
\subfigure[Splitting by state (without core)]{
\includegraphics[width=0.31\hsize]{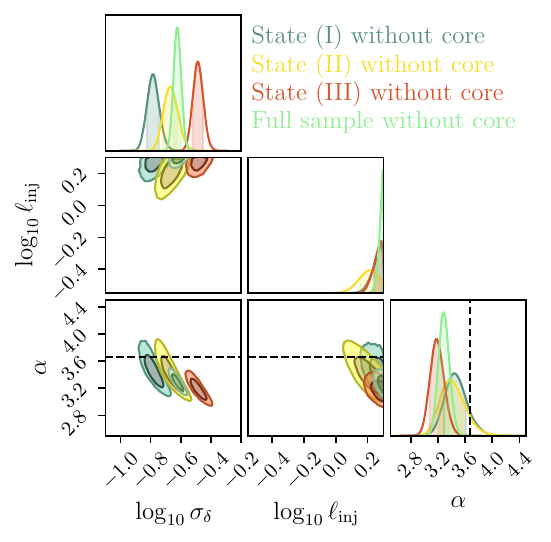}
\label{fig:joint_state_without_center}
}
\subfigure[Splitting by mass (without core)]{
\includegraphics[height=0.31\hsize]{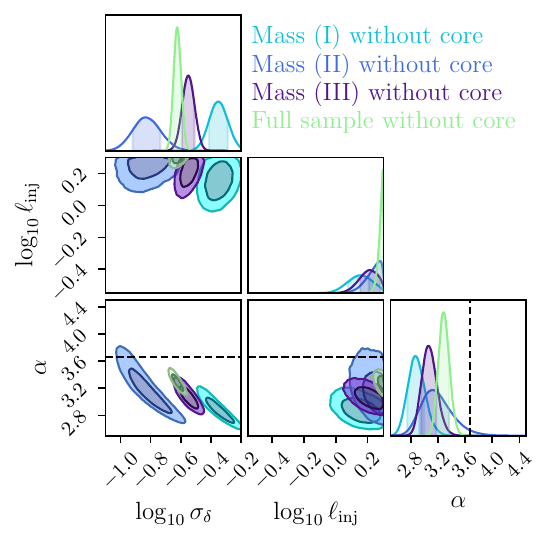}
\label{fig:joint_mass_without_center}
}
\subfigure[Splitting by redshift (without core)]{
\includegraphics[width=0.31\hsize]{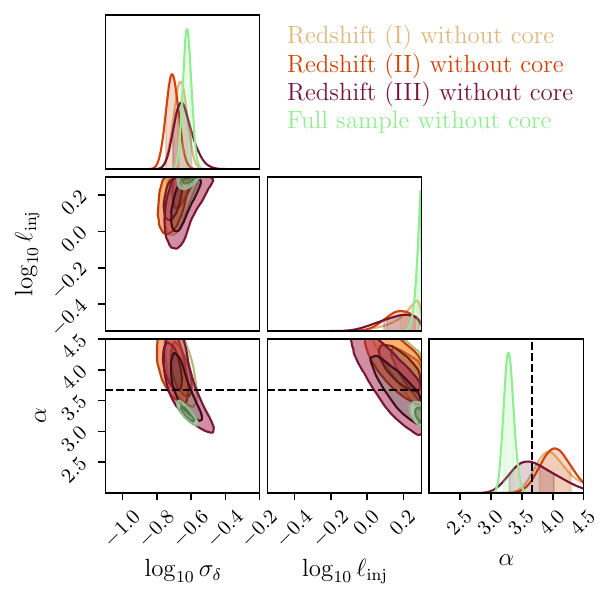}
\label{fig:joint_z_without_center}
}
\caption{(a) Distribution of the subsamples in the mass-redshift plane. The shaded areas indicate the mass and redshift subsamples extracted from Tiers 1 and 2 respectively. The colour of the dots reflects the dynamical state of the clusters as measured by the centroid shift $w$. (b) Joint posterior distributions of the standard deviation, $\sigma_\delta$, the injection scale, $\ell_\text{inj}$, and the spectral index $\alpha$ of the density fluctuation power spectrum parameters, evaluated on the full sample including the core $< 0.15 R_{500}$ region. The black dashed line represents the expected 11/3 index from Kolmogorov theory. (d), (e) and (f) are the same plot using the subsamples associated respectively with the state, mass, and redshift as defined in Tab.~\ref{tab:subsamples_definition}, including the core region. (g), (h), and (i) use the same subsamples without the core regions.}
\label{fig:joint_corners}
\end{figure*}

The posterior distributions obtained for the joint fit across the state, mass, and redshift subsamples, as defined in Tab.~\ref{tab:subsamples_definition}, are illustrated in Fig.~\ref{fig:with_core} when the core is included and in Fig.~\ref{fig:without_core} where the core is excluded. The separation of the dynamical states (Fig.~\ref{fig:joint_state} and Fig.~\ref{fig:joint_state_without_center}) shows an increase in the variance of the fluctuations with the centroid shift $w$. These results can also be compared with those found for the X-COP sample \citep{dupourque_investigating_2023}, which showed a positive correlation between dynamical indicators that increase with cluster perturbation, specifically the centroid shift $w$ and the Zernike moments. Despite such differences, a Kolmogorov/hydrodynamical cascade (with -11/3 slopes) appears to be a good approximation for the type of turbulence developing in such systems, { as the slope is constrained to  $\alpha = 3.58 \pm 0.11$ including the core and $\alpha = 3.28 \pm 0.08$ excluding it. The flatter slope in the outer regions might be correlated with a more predominant magnetic field \citep{vestuto_spectral_2003}.}

The separation of the Tier 1 sample into several mass bins (Fig.~\ref{fig:joint_mass} and Fig.~\ref{fig:joint_mass_without_center}) shows a non-trivial correlation between mass and variance of the fluctuations. Regarding previous observational constraints, \citet{hofmann_thermodynamic_2016} find a mild anti-correlation between relative density fluctuations (or Mach number) and masses, decreasing by a factor of 1.5$\times$ from poor to massive clusters. This is in line with our highest and lowest mass bins but not with the intermediate mass bin which shows much lower normalisations. Note that the low-mass bin does not contain any bin State (I) objects, i.e. the most relaxed clusters in this sample, the high level of normalisation measured can therefore also be linked to the dynamical state within the bin. As for the injection scale, clusters are expected to virialize self-similarly, however, meso- and micro-scale astrophysics introduces key deviations depending on the hydrodynamical process at play (e.g., feedback, condensation, sloshing). For instance, the injection scale tends to reside near the $R_{500}$ region, but lower mass systems are capable to reach smaller injection scales toward the core region, likely due to the relatively stronger AGN feedback impact (e.g. \citealt{gaspari_can_2014}). The slope is instead roughly invariant with mass, again suggesting a similar type of turbulence (hydrodynamical). The Mass (III) sample is the most comparable to the previous study on X-COP, and shows comparable variance and slope, but increased injection, which is interpreted as a spatial resolution effect in Sec.~\ref{sec:core-exclusion}.

The last corner plots (Fig.~\ref{fig:joint_z} and Fig.~\ref{fig:joint_z_without_center}) show the separation of the sample over the Tier 2 redshift range. At high redshifts $(z\gtrsim1)$, galaxy clusters are expected to experience more frequent mergers \citep[e.g.][]{husko_statistics_2022}, which can induce stronger turbulence and density fluctuations in the ICM. With increasing redshift, we observe a mild increase in density variance (hence turbulence) and injection scales. However, our redshifts reach at best up to $z\sim0.5-0.6$, far from the early universe formation. Thus, over the redshift range studied via Tier 2, most turbulence parameters appear to be fairly similar. 

\subsection{Effect of core exclusion \& spatial resolution}
\label{sec:core-exclusion}

The first differences between the analyses with and without the cluster core involve an increase in fluctuation variance, especially in the Mass (III), State (III) and global results, which might suggest that clusters are more disturbed on average in the outer regions. There is also a decrease in the global slope, which could indicate either that the outer regions are noisier due to a lower signal-to-noise ratio or that turbulent processes undergo reinjection at multiple scales, thereby flattening the observed spectrum. About the injection scale, energy injection processes in the cluster cores, such as AGN feedback and sloshing, have a smaller characteristic scale than the outer regions, which are dominated by accretion and merger events. Our constraints on the injection scale average the low injections in the centre and the high injections from the outer parts. This might explain the reason for which we constrain $\ell_{\text{inj}}$ to approximately $R_{500}$ when including the core region but reduce it to an upper limit of roughly $2 R_{500}$—the largest scale that can be measured in our field—when the core is excluded. Using CHEX-MATE, the low angular size of clusters prevents us from assessing properly the low injection scale in the central region, due to the low number of pixels in this region. As an illustration, there is a median number of $\sim 200$ pixels below 0.15 $R_{500}$ in CHEX-MATE, compared to $\sim 10 000$ pixels below 0.15 $R_{500}$ in X-COP. However, such behaviour has already been observed in the radial study of X-COP. 

Furthermore, as CHEX-MATE clusters are, on average, smaller in the sky than X-COP clusters, the lowest spatial limit that can be exploited on X-COP is 50\% smaller than that on CHEX-MATE, which implies a better sampling of the inertial range. This is illustrated in Fig.~\ref{fig:low_scale_redshift}, where the conservative lowest scale we defined is shown along the individual scales that could be achieved for each cluster, and the clusters common with X-COP are on the lower left. This definition for the lowest scale allows consistency of analysis over the whole sample but results in less sensitivity to small scales which are found in the core regions of clusters, skewing the results toward larger injection scales. In Fig.\ref{fig:joint_with_xcop}, the constraints from X-COP align with those from CHEX-MATE when we compare the outer rings, which exclude the core region driving the low injection. Systematically excluding the core from this kind of analysis will reduce the contamination from substructures such as the sloshing spirals, the cool core or cold fronts from older merger, which further increase the correlation between measured density fluctuations and turbulent process, see Sec.~\ref{sec:correlation_delta_m} for further discussions.

\subsection{Correlation with radio data}
\label{sec:radio-halo}
The presence of turbulence in the ICM is expected to induce particle acceleration through second order Fermi mechanisms in turbulent medium \citep[e.g.][]{brunetti_stochastic_2016} which could be responsible for the presence of radio haloes in galaxy clusters. \cite{eckert_connection_2017} showed a bimodality in the distribution of the amplitude of X-SB fluctuations at 150 kpc, separating clusters with and without radio haloes and the correlation between the radio halo power at 1.4 GHz and the one dimensional Mach number $\mathcal{M}_{1D}$, using a sample of 51 clusters. More recently, \cite{zhang_planck_2023} performed a similar analysis on a smaller sample of 36 halos, detected at lower frequencies with LOFAR. Confirmation of the bimodality was not achievable due to the limited sample size. Additionally, no clear indications of a correlation between the velocity dispersion and the radio power at 150 MHz were found, possibly due to the LoTSS DR2 survey containing more objects with an ultra-steep spectrum. These objects experienced less energetic mergers when compared to clusters in \cite{eckert_connection_2017} (see \cite{brunetti_cosmic_2014} for a review). In this section, we use the CHEX-MATE, which is partially covered by the LoTSS \citep{shimwell_lofar_2019}, searching for a trend in our density fluctuation model regarding the presence of radio halo, using the whole available spatial scales. 

\begin{figure*}[t]
    \centering
    \subfigure{
\includegraphics[height=0.45\hsize]{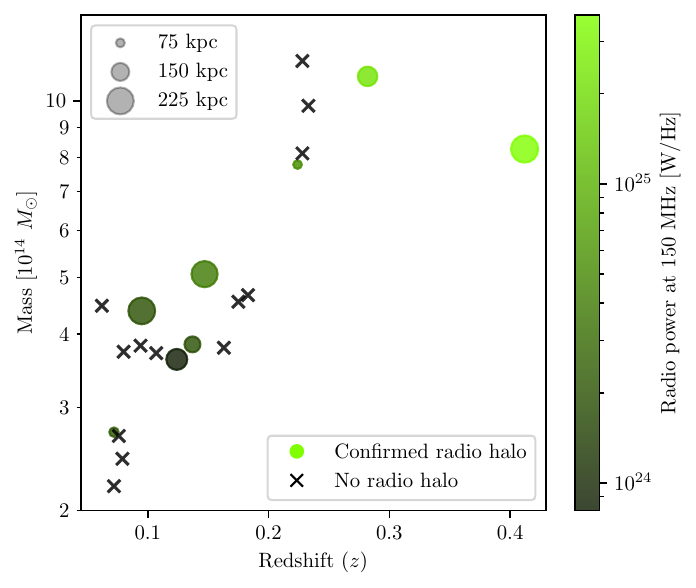}}
\subfigure{
\includegraphics[width=0.45\hsize]{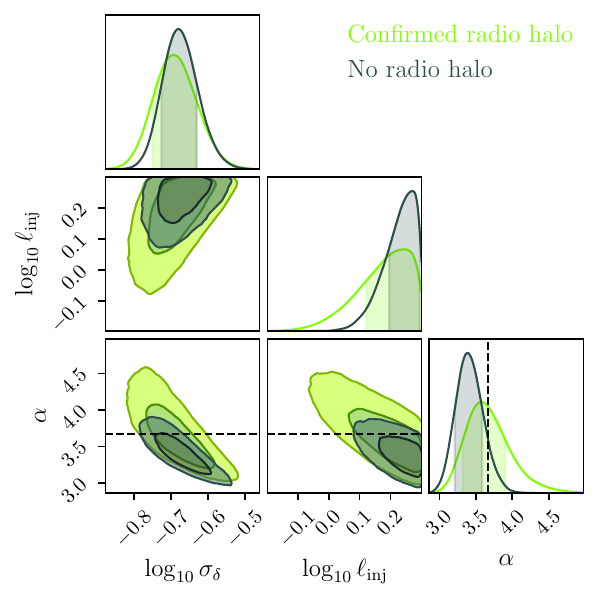}}
    \caption{(\textit{left}) Visualization of the clusters in LoTSS DR2 in the mass-redshift plane, with estimated halo size and power at 150 MHz. (\textit{right}) Density fluctuation parameters estimated when using subsamples with confirmed or excluded radio halo.}
    \label{fig:radio_contour}
\end{figure*}

In Fig.~\ref{fig:radio_contour}, we plot the contour obtained for the analysis split into the common clusters with LoTSS DR2, with and without radio halo. We note that, on average, the parameters we examine align well. We do not observe any significant difference between the two subsamples. It is important to note that the exclusion of the most perturbed clusters leads us to study a population of radio-halo with a homogeneous distribution regarding their dynamical state, as illustrated in Fig.~\ref{fig:mixing-between-sates}. Hence, this population is very different from the populations previously studied in \cite{eckert_connection_2017} and \cite{zhang_planck_2023}, which may justify the absence of a clear trend in our results. The limited sample size hinders the determination of trends between fluctuations and radio halos in this study, but the upcoming radio coverage of the CHEX-MATE sample (82 clusters in total) will allow for the confirmation or refutation of this trend.

\section{Discussions}
\label{sec:discussions}

\subsection{Interpretation as turbulent motion }

Assuming turbulence in the ICM as the primary source of relative density fluctuations, precursory hydrodynamical and theoretical studies (\citealt{gaspari_constraining_2013, gaspari_relation_2014, zhuravleva_relation_2014}) demonstrated that the characteristic amplitude of turbulent velocity dispersion -- which defines the Mach number $\mathcal{M} = \sigma_v/c_{\rm s}$ (with $\sigma_v$ the 3D velocity dispersion and $c_{\rm s}$ the sound speed) -- is \textit{linearly} tied to the characteristic amplitude of density fluctuations, unlike in non-stratified fluids, where a quadratic trend is expected. This relation has also been studied in the context of turbulence in a box and corroborated for major stratification levels \citep{mohapatra_turbulence_2019, shi_turbulence_2019, mohapatra_turbulence_2020, mohapatra_turbulent_2021}. \cite{gaspari_constraining_2013} initially found the linear scaling relation between the density fluctuations at different scales and the associated Mach number, by testing varying plasma physics (turbulence, thermal conduction, electron-ion equilibration) in high-resolution hydrodynamical simulations of galaxy clusters. As thermal conduction is expected to be highly suppressed in the observed ICM, an argument supported by our nearly Kolmogorov slopes and by theoretical work \citep[e.g.][]{zuhone_cold_2016}, the relation from \cite{gaspari_constraining_2013}, involving the peak of the amplitude spectrum, can be simplified to :

\begin{equation}
\mathcal{M}_{3D} \approx 2 \sigma_{\delta} \left(\frac{l_\text{inj}}{L_\text{500}}\right)^{-1/5},
\label{eq:g13}
\end{equation}
where $L_\text{500}$ is an injection scale of 500 kpc. Accounting for different levels of turbulence and related spectral slopes, this adds a scatter of $\sim$\,0.1\,dex.

\begin{figure}[t]
    \centering
\includegraphics[width=\hsize]{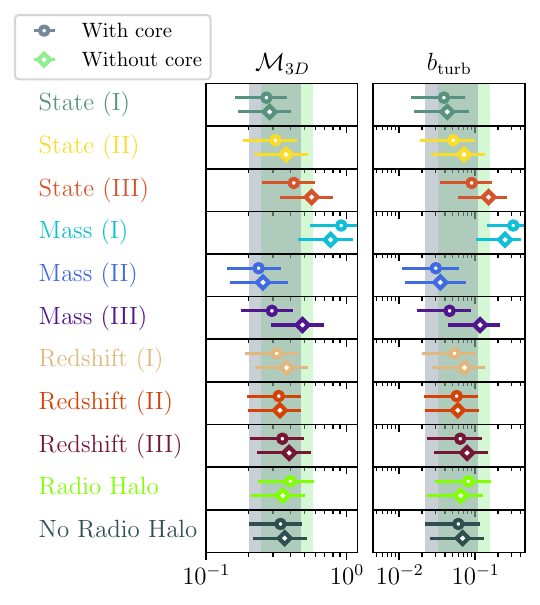}
    \caption{Mach number $\mathcal{M}_{3D}$ and turbulent mass bias $b_{\text{turb}}$ as obtained using the relationship from Eq.~\ref{eq:yet-another-scaling} and Eq.~\ref{eq:non-thermal-ratio} respectively, applied to the dynamical state, the mass, the redshift and the radio halo subsamples. The blue and green bands represent the average value over the whole sample, respectively, with and without the core. (I) to (III) relates to the increasing value which has been used to define the subsample, i.e. increasing disturbance for the state, and increasing mass and redshift. See Tab.~\ref{tab:subsamples_definition} for proper definition of each subsample.}
    \label{fig:mach_nth_support}
\end{figure}

Subsequently, \cite{zhuravleva_turbulent_2014} and \cite{gaspari_relation_2014} retrieved an analogous linear relationship with the one-dimensional Mach number ($\mathcal{M}_{1D}=\mathcal{M}_{3D}/\sqrt{3}$):
\begin{equation}
    \mathcal{M}_{1D} \approx (1 \pm 0.3)\, \sigma_\delta.
    \label{eq:z14-g14}
\end{equation}
Overall, the above-quoted studies converge toward a linear relation with roughly unitary normalization with $\mathcal{M}_{1D}$. Recently, this was also validated in a set of cosmological simulations with a sample of 80 clusters (\citealt{zhuravleva_indirect_2023}), in particular for non-merging systems, which is also the focus of our work. Eq.~\ref{eq:g13} has a low dependency on the injection scale, and is comparable to the scaling from Eq.~\ref{eq:z14-g14}. To account for the dispersion between the two relations, we choose to convert the density fluctuations in Mach number using a unity scaling with a dispersion that accounts for the dispersion between Eq.~\ref{eq:g13} and Eq.~\ref{eq:z14-g14}:

\begin{equation}
    \mathcal{M}_{3D} \approx \sqrt{3} \times (1 \pm 0.4)\, \sigma_\delta.
    \label{eq:yet-another-scaling}
\end{equation}

Further, we can estimate the contribution of such processes to the non-thermal pressure support globally occurring in our sample. The ratio between the non-thermal pressure and the total pressure can be expressed as a function of the 3D turbulent Mach number $\mathcal{M}_{3D}$ \citep{eckert_non-thermal_2019} if the non-thermal contribution exclusively comes from turbulence:

\begin{equation}
    \frac{P_{\text{turb}}}{P_{\text{tot}}} = \frac{\mathcal{M}_{3D}^2\gamma}{\mathcal{M}_{3D}^2\gamma + 3},
    \label{eq:non-thermal-ratio}
\end{equation}
where $\gamma = 5/3$ is the polytropic index. The ratio from Eq.~\ref{eq:non-thermal-ratio} can directly be interpreted as a turbulent mass bias $b_{\text{turb}}$ if $\gamma = 5/3$. The Mach numbers and turbulent mass bias $b_{\text{turb}}$ deduced in $R_{500}$ are shown in Fig.~\ref{fig:mach_nth_support} and Tab.~\ref{tab:mach-number} and Tab.~\ref{tab:mach-number-without-center} for each subsamples, using the scaling from Eq.~\ref{eq:yet-another-scaling}, with a $\mathcal{M}_{3D} \simeq 0.41 \pm 0.17$ when evaluated on the whole sample without the core. Consistently with previous studies (e.g., \citealt{hofmann_thermodynamic_2016}) and physical expectations, 3D Mach numbers remain in the subsonic regime, typically in the 0.3\,-\,0.5 range. { Direct comparison with existing works is not straightforward, as the value of the $\mathcal{M}_{3D}$ derived in this work requires comparison to the 1D velocity broadening measurements. Moreover, the analysis regions we used cover a much larger fraction of the cluster volume than the core region that is probed in most other works. In any case, our results are compatible with subsonic motions as implied by \cite{sanders_constraints_2011, sanders_velocity_2013, pinto_chemical_2015}} The measured $\mathcal{M}_{3D}$ number value translates into turbulent pressure support of $P_{\text{turb}}/P_{\text{tot}} \simeq (9\pm 6) \%$, which is in line with the typical value retrieved in cosmological simulations (e.g., \citealt{vazza_turbulent_2018}), and previous observational works on the X-COP sample \citep{eckert_non-thermal_2019, dupourque_investigating_2023}. The bias of $b = 0.09\pm0.06$ the entire considered sample is compatible with the value obtained in \cite{lovisari_chex-mate_2023} when accounting for the fluctuations at all scales, and previous observational works in X-ray \citep[e.g.][]{eckert_non-thermal_2019, ettori_tracing_2022} or using lensing \citep[e.g.][]{mahdavi_evidence_2008, zhang_locuss_2010}. As noted in Sec.~\ref{sec:constraints}, Fig.~\ref{fig:mach_nth_support} also shows a clear increase in turbulent Mach numbers toward more disturbed systems, as well as toward lower cluster masses. The redshift trend is instead minor, as expected by our limited range $z< 0.6$. A complementary approach, proposed by \cite{lovisari_chex-mate_2023}, determines the Mach number and the non-thermal contribution from temperature fluctuations measured on a CHEX-MATE subsample. By performing our joint analysis on the same subsample, we obtain a Mach number of the order of $\mathcal{M}_{3D} \simeq 0.41 \pm 0.18$ which is compatible with $\mathcal{M}_{3D} \simeq 0.36\pm 0.10$ as obtained from \cite{lovisari_chex-mate_2023}. We note that the resolution of the X-SB maps is better than the resolution of the temperature maps, so the most sensitive scales of turbulence might be different. { We stress that these results are derived under the assumption of homogeneous and isotropic turbulence, which consequently is described by a Gaussian random field. The stratification of the turbulence, its multiphase aspect, and its general anisotropies constitute departures from the assumption of Gaussianity. In particular, not taking stratification into account \citep{mohapatra_turbulent_2021} and not considering accreted clumps or structural merger remnants may lead to an overestimate of our turbulent Mach number. Accounting for the stochasticity of this process via the sample variance constitutes a non-negligible barrier when it comes to accurately constrain more complex models. Nevertheless, our use of a sample of clusters and the exclusion of central regions limits contamination by clumps, and our use of scaling relations derived from cosmological simulations \citep{gaspari_constraining_2013, simonte_exploring_2022, zhuravleva_indirect_2023} partially considers of the effect of anisotropies into our final Mach number calculation.}

\subsection{Discussing the correlation between $\sigma_\delta$ and $\mathcal{M}_{3D}$}
\label{sec:correlation_delta_m}

The linear link between density fluctuations and turbulent velocities can be analytically derived via simplified hydrodynamics equations for a stratified medium like the ICM \citep{gaspari_constraining_2013, zhuravleva_relation_2014}. This correlation has been investigated and tested using hydrodynamical simulations \citep{gaspari_constraining_2013, zhuravleva_relation_2014, gaspari_relation_2014, simonte_exploring_2022, zhuravleva_indirect_2023}. 
At variance with the above works, \cite{simonte_exploring_2022} applied small-scale filtering techniques to disentangle the spurious contribution of bulk flows and extract the isotropic part of turbulent motions, and excised gas clumps in the simulated ICM. Following this approach, they reported a substantially lower normalization (factor of 2-to-3) and larger scatter in the above relation, also with significant differences in the slope of the relation, depending on the dynamical state of clusters.  They concluded that, in general, the presence of residual gas density fluctuations not directly linked to turbulence, but to the crossing of self-gravitating clumps and substructures in the ICM, is hard to excise. It introduces a large scatter in the observational link between projected density fluctuations and the true turbulent budget of the ICM.  On the other hand, without distinguishing between turbulent and bulk gas motions in the ICM, cosmological runs by \cite{zhuravleva_indirect_2023} show such mild/strong correlation and normalization. To disentangle both sources of fluctuations, a clump filtering method has to be applied to simulated and observed datasets. Typically, this involves identifying the high-density tail of the gas density distribution. For example, \cite{simonte_exploring_2022} cut the top 5\% tail, while \cite{zhuravleva_indirect_2023} more conservatively cut less than 0.05\% of the values (3.5\,$\sigma$). The former reports an applied threshold $\sigma_\delta^2 = 0.22$, hence potentially filtering motions $\mathcal{M}_{1D} \ge 0.47$, which may account for the difference among the various theoretical predictions. 

From an observational perspective, it is non-trivial to understand the nature of gas clumping in clusters. Besides filtering obvious mergers, the results of our analysis can be biased toward some residual density structures, which may have varying effects on the parameters depending on their distance from the centre of the cluster, its mass and its dynamical state. Gas clumps in the outer parts of the cluster will induce high $\sigma_\delta$ as they represent a large relative fluctuation, which could contribute to the correlation between dynamical states and fluctuation normalizations as disturbed clusters contain more substructures. In the central parts, gas clumps are understood as highly significant fluctuations in a very dense region, which will drive the overall spectrum to lower values of $\sigma_\delta$. The highly structured fluctuations associated with clumps can also bring high slopes and constraints on the injection scale that are close to the clump size. This highlights the challenge of disentangling density fluctuations emerging from turbulent motions and substructures/clumps, such as those also observed in the central parts of X-COP clusters \citep{dupourque_investigating_2023} and in the CHEX-MATE sample, as illustrated in Fig.~\ref{fig:gallery}. One approach to estimating the contamination due to these substructures could be to use simulations involving contamination by modelled cool-cores or sloshing spirals, and to estimate the biases induced on the reconstruction of the parameters of the fluctuation field as well as the robustness of this approach to the individual deviations of each cluster. This exercise would require the exploration of a large space of contaminant parameters, and will be the subject of a future study. Nevertheless, the results obtained by excluding the central regions of each cluster should be more robust to the presence of artifacts, as shown by the exclusion of the bullet of PSZ2G266.04-21.25, and other clumps in Fig.~\ref{fig:gallery}.
%-----------------------------------------------------------------

\section{Conclusions}

We have applied the methodology introduced in \cite{dupourque_investigating_2023} to the CHEX-MATE sample. We focused on a sample of 64 objects, excluding the most disturbed clusters, as our analysis is not suitable for strongly perturbed surface brightness images. We used this subsample to characterise the gas density fluctuations through simulation-based inference applied to the power spectrum of surface brightness fluctuations, increasing the statistic by a factor of $\sim 5$ when compared to X-COP.

\begin{itemize}
    \item We derived the parameters of the density fluctuation spectrum for different CHEX-MATE subsamples. The high sample statistic allows us to define 3 classes in terms of dynamical state, mass, and redshift. The fluctuations show a normalisation that increases with the dynamical state, which relates to the fact that disturbed clusters tend to have more density fluctuations. We observe an anticorrelation between mass and normalisation of the fluctuations. The redshift subsamples show invariant behaviour, which is consistent with the fact that the turbulence within the ICM is not expected to change drastically over the studied redshift range ($z<0.6$).
    \item We assessed the effects of cluster core on the overall results by comparing the analysis in $R_{500}$ to the analysis excluding the central $0.15 R_{500}$ in each cluster. It can be seen that, in particular, the injection scale is no longer constrained when the centre is excluded, which means that the injection we measure is mainly driven by the high signal and low scale processes occurring in the central region, while the signal outside $0.15 R_{500}$ only allows us to establish lower limits. The core exclusion reduces the contamination from central structures such as the cool-cores or sloshing spiral, making a more reliable measurement for the statistics of fluctuations. 
    \item We conducted a correlation analysis between our measurements and radio data from the LoTSS-DR2, investigating potential relationships between density fluctuation parameters and the presence of radio halos. No specific trends emerged when comparing individual parameters of clusters with associated radio halo parameters, which is in line with expectations from the population of haloes we studied, which are low-frequency radio haloes in relaxed clusters.
    \item We interpreted the density fluctuations as resulting from turbulent processes. Using two scaling relations, we derived 3D Mach numbers for various subsamples defined using the CHEX-MATE sample. We obtained an average $\mathcal{M}_{3D} \simeq 0.4 \pm 0.2$ and a corresponding non-thermal pressure support of $ P_{\text{turb}}/P_{\text{tot}} \simeq 9\pm 6 \%$ or $b_{\text{turb}} \simeq 0.09 \pm 0.06$ in $R_{500}$, which is consistent with what can be found in the literature.
\end{itemize}

More advanced constraints could be obtained in the future by crossing the fluctuations of the different ICM observables, such as the fluctuations in the SZE, which carry information at further distances from the centre thanks to dependency over $n_e$ instead of $n_e^2$ for the X-Ray emissivity. Pioneering work such as \cite{khatri_thermal_2016} and \cite{romero_inferences_2023} demonstrated the feasibility of this approach, yet it remains very challenging even using high resolution SZ instruments and still has to integrate the sample variance. In addition, direct and unambiguous measurement of gas motions could be possible in the coming year with \textit{XRISM}/Resolve \citep{terada_detailed_2021} and in the next decade with the Line Emission Mapper \citep[LEM][]{kraft_line_2022} and \textit{Athena}/X-IFU \citep{barret_athena_2020}, through spatially resolved observations of spectral lines centroid shift and broadening.

\begin{acknowledgements}
EP, NC, SD, and GWP acknowledge the support of CNRS/INSU and CNES. LL acknowledges the financial contribution from the INAF grant 1.05.12.04.01. MDP acknowledges support from Sapienza Università di Roma thanks to Progetti di Ricerca Medi 2020, RM120172B32D5BE2 and Medi 2021, RM12117A51D5269B. FV acknowledges the financial support from the Cariplo "BREAKTHRU" funds Rif: 2022-2088 CUP J33C22004310003. MR, IB, SE, SG acknowledge the financial contribution from the contracts Prin-MUR 2022, supported by Next Generation EU (n.20227RNLY3 {\it The concordance cosmological model: stress-tests with galaxy clusters}), ASI-INAF Athena 2019-27-HH.0,``Attivit\`a di Studio per la comunit\`a scientifica di Astrofisica delle Alte Energie e Fisica Astroparticellare'' (Accordo Attuativo ASI-INAF n. 2017-14-H.0), and from the European Union’s Horizon 2020 Programme under the AHEAD2020 project (grant agreement n. 871158). This work was granted access to the HPC resources of CALMIP supercomputing center under the allocation 2022-22052. LOFAR data products were provided by the LOFAR Surveys Key Science project (LSKSP; \url{https://lofar-surveys.org/}) and were derived from observations with the International LOFAR Telescope (ILT). LOFAR \citep{haarlem_lofar_2013} is the Low Frequency Array designed and constructed by ASTRON. It has observing, data processing, and data storage facilities in several countries, that are owned by various parties (each with their own funding sources), and that are collectively operated by the ILT foundation under a joint scientific policy. The efforts of the LSKSP have benefited from funding from the European Research Council, NOVA, NWO, CNRS-INSU, the SURF Co-operative, the UK Science and Technology Funding Council and the Jülich Supercomputing Centre. This research was supported by the International Space Science Institute (ISSI) in Bern, through ISSI International Team project \#565 ({\it Multi-Wavelength Studies of the Culmination of Structure Formation in the Universe}).
This work made use of various open-source packages such as 
\texttt{matplotlib} \citep{hunter_matplotlib_2007},
\texttt{astropy} \citep{robitaille_astropy_2013, the_astropy_collaboration_astropy_2018},
\texttt{ChainConsumer} \citep{hinton_chainconsumer_2016},
\texttt{cmasher} \citep{velden_cmasher_2020},
\texttt{sbi} \citep{tejero-cantero_sbi_2020},
\texttt{pyro} \citep{bingham_pyro_2019},
\texttt{jax} \citep{bradbury_jax_2018},
\texttt{haiku} \citep{hennigan_haiku_2020},
\texttt{numpyro} \citep{bingham_pyro_2019, phan_composable_2019}

\end{acknowledgements}


\bibliographystyle{aa} % style aa.bst
\bibliography{references.bib}

\begin{appendix} %First appendix

\section{Fourier transform convention}
\label{app:fourier_convention}

In this paper, we define the Fourier transform with the classical signal processing convention, namely $(0, -2\pi)$, see \cite{weisstein_fourier_1995}. This pairs results in the forward transform highlighted in Eq. \ref{eq:tf2D} and \ref{eq:tf3D}. We use $\hat{f}$ and $\tilde{f}$ to refer respectively to the 2D and 3D Fourier transform of a function $f$.

\begin{equation}
\label{eq:tf2D}
    \mathcal{FT}_{2D}\left\{f\right\} \equiv \int \diff^2 \Vec{\rho} \, f(\Vec{\rho}) e^{ -2i\pi\Vec{k_\rho}.\Vec{\rho}} =  \hat{f}(\vec{k}_\rho)
\end{equation}

\begin{equation}
\label{eq:tf3D}
    \mathcal{FT}_{3D}\left\{f\right\} \equiv \int \diff^3\Vec{r} \, f(\vec{r}) e^{ -2i\pi\Vec{k_r}.\Vec{r}} =  \tilde{f}(\vec{k}_r)
\end{equation}

%\section{Excluded clusters}

\section{Table with values}
\label{app:table_with_values}

The results obtained for the variance of density fluctuations $\sigma_{\delta}$, the injection scale $\ell_{\text{inj}}$, the spectral index $\alpha$, the 3D Mach number $\mathcal{M}_{3D}$ and the turbulent mass bias $b_{\text{turb}}$ are displayed in Tab.~\ref{tab:mach-number} with the core region is included and in Tab.~\ref{tab:mach-number-without-center} with the 0.15 $R_{500}$ inner region excluded.

\begin{table*}[t]
    \centering
    \begin{tabular}{l|c|c|c|c|c}
        Subsample & $\sigma_{\delta}$ & $\ell_{\text{inj}}$ & $\alpha$ & $\mathcal{M}_{3D}$ & $b_{\text{turb}}$\\
        \hline
        \hline
State (I) & $0.16\pm 0.01$&$1.03\pm 0.21$&$3.34\pm 0.15$&$0.27\pm 0.11$&$0.04\pm 0.03$\\
State (II) & $0.18\pm 0.02$&$1.45\pm 0.23$&$3.24\pm 0.12$&$0.31\pm 0.13$&$0.05\pm 0.04$\\
State (III) & $0.24\pm 0.02$&$0.87\pm 0.24$&$3.71\pm 0.29$&$0.42\pm 0.17$&$0.09\pm 0.07$\\
Mass (I) & $0.53\pm 0.06$&$1.62\pm 0.27$&$2.64\pm 0.09$&$0.91\pm 0.39$&$0.32\pm 0.16$\\
Mass (II) & $0.14\pm 0.02$&$0.60\pm 0.20$&$3.30\pm 0.25$&$0.24\pm 0.10$&$0.03\pm 0.03$\\
Mass (III) & $0.17\pm 0.01$&$0.73\pm 0.16$&$3.85\pm 0.18$&$0.29\pm 0.12$&$0.05\pm 0.04$\\
Redshift (I) & $0.18\pm 0.02$&$0.94\pm 0.25$&$3.94\pm 0.23$&$0.32\pm 0.13$&$0.05\pm 0.04$\\
Redshift (II) & $0.19\pm 0.02$&$1.30\pm 0.31$&$3.56\pm 0.27$&$0.33\pm 0.14$&$0.06\pm 0.05$\\
Redshift (III) & $0.20\pm 0.03$&$1.12\pm 0.27$&$3.57\pm 0.29$&$0.35\pm 0.15$&$0.06\pm 0.05$\\
Radio Halo & $0.24\pm 0.04$&$1.48\pm 0.53$&$3.47\pm 0.47$&$0.40\pm 0.18$&$0.08\pm 0.07$\\
No Radio Halo & $0.20\pm 0.02$&$0.84\pm 0.24$&$3.25\pm 0.19$&$0.34\pm 0.14$&$0.06\pm 0.05$\\
Joint & $0.19\pm 0.01$&$1.04\pm 0.14$&$3.58\pm 0.11$&$0.34\pm 0.14$&$0.06\pm 0.04$\\

    \end{tabular}
    \caption{Joint marginalized constraints on the parameters for each of the subsamples defined in Tab.~\ref{tab:subsamples_definition}, uncertainties correspond to the 68\% confidence ranges.}
    \label{tab:mach-number}
\end{table*}

\begin{table*}[t]
    \centering
    \begin{tabular}{l|c|c|c|c|c}
        Subsample & $\sigma_{\delta}$ & $\ell_{\text{inj}}$ & $\alpha$ & $\mathcal{M}_{3D}$ & $b_{\text{turb}}$\\
        \hline
        \hline
State (I) & $0.16\pm 0.02$&$1.83\pm 0.14$&$3.46\pm 0.16$&$0.28\pm 0.12$&$0.04\pm 0.04$\\
State (II) & $0.22\pm 0.02$&$1.60\pm 0.21$&$3.41\pm 0.18$&$0.37\pm 0.16$&$0.07\pm 0.06$\\
State (III) & $0.33\pm 0.03$&$1.84\pm 0.12$&$3.19\pm 0.10$&$0.57\pm 0.23$&$0.15\pm 0.10$\\
Mass (I) & $0.45\pm 0.06$&$1.42\pm 0.24$&$2.88\pm 0.13$&$0.77\pm 0.34$&$0.25\pm 0.14$\\
Mass (II) & $0.15\pm 0.03$&$1.75\pm 0.19$&$3.16\pm 0.23$&$0.25\pm 0.12$&$0.03\pm 0.04$\\
Mass (III) & $0.28\pm 0.03$&$1.61\pm 0.21$&$3.07\pm 0.11$&$0.49\pm 0.20$&$0.12\pm 0.08$\\
Redshift (I) & $0.22\pm 0.02$&$1.71\pm 0.26$&$4.00\pm 0.31$&$0.37\pm 0.16$&$0.07\pm 0.05$\\
Redshift (II) & $0.19\pm 0.02$&$1.50\pm 0.24$&$4.05\pm 0.25$&$0.34\pm 0.14$&$0.06\pm 0.05$\\
Redshift (III) & $0.22\pm 0.03$&$1.47\pm 0.30$&$3.70\pm 0.35$&$0.39\pm 0.17$&$0.08\pm 0.06$\\
Radio Halo & $0.20\pm 0.03$&$1.55\pm 0.28$&$3.64\pm 0.31$&$0.35\pm 0.15$&$0.06\pm 0.05$\\
No Radio Halo & $0.21\pm 0.02$&$1.72\pm 0.20$&$3.41\pm 0.19$&$0.37\pm 0.16$&$0.07\pm 0.05$\\
Joint & $0.24\pm 0.01$&$1.93\pm 0.07$&$3.29\pm 0.08$&$0.41\pm 0.17$&$0.09\pm 0.06$\\

    \end{tabular}
    \caption{Joint marginalized constraints on the parameters for each of the subsamples defined in Tab.~\ref{tab:subsamples_definition}, after exclusion of the central 0.15 R500 region. Uncertainties correspond to the 68\% confidence ranges.}
    \label{tab:mach-number-without-center}
\end{table*}

\section{Fluctuation gallery}

\begin{figure*}[t]
    \centering

\includegraphics[height=0.95\vsize]{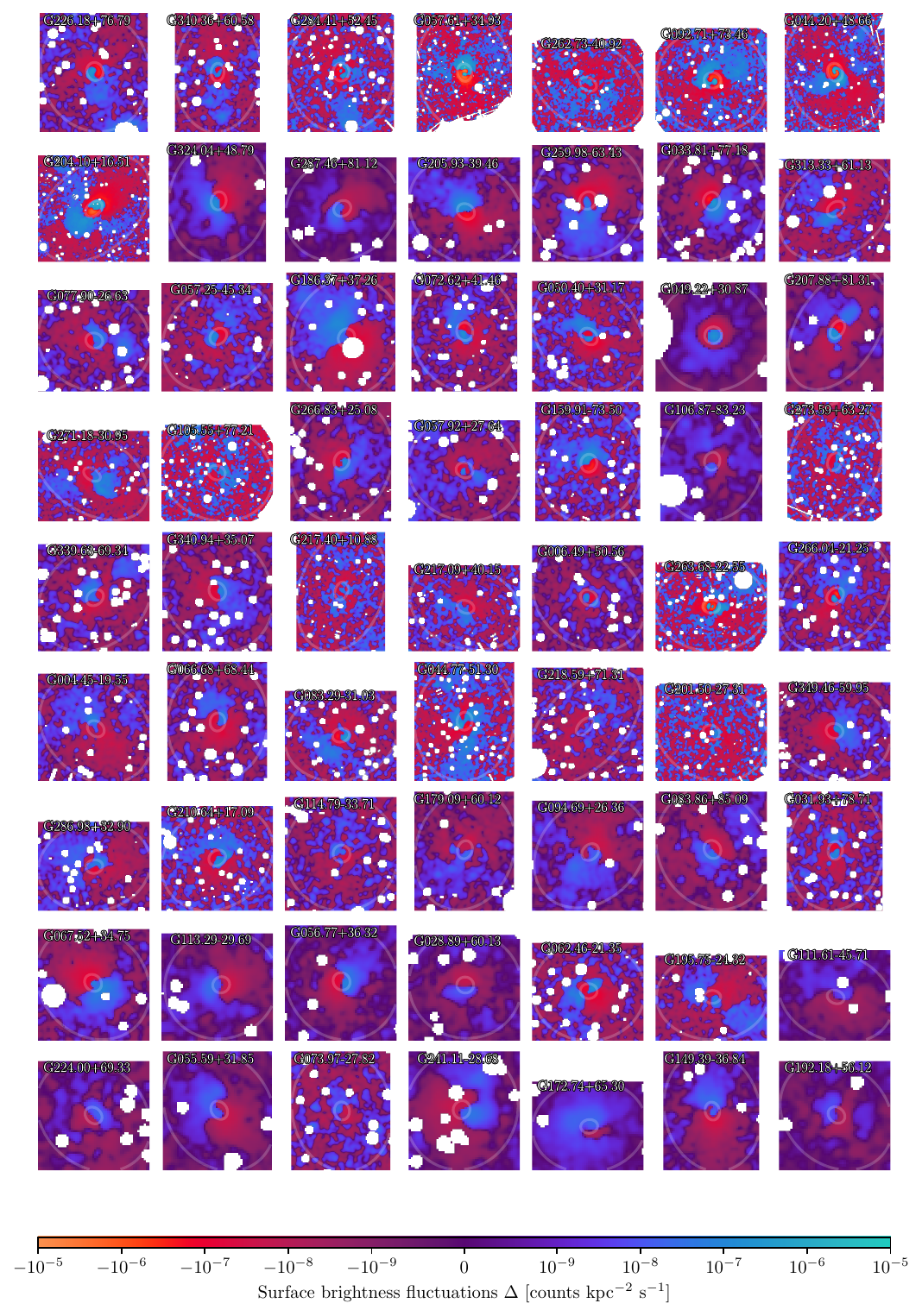}
    \caption{Surface brightness fluctuation gallery with masked point sources, as defined in Eq.~\ref{eq:sb-fluc-diff} for the 63 clusters used in the analysis (see Fig.~\ref{fig:subsamples_definition}), in ascending order of $w$. The two rings denote $0.15$ and $1 R_{500}$.}
    \label{fig:gallery}
\end{figure*}

The gallery of surface brightness fluctuations on the CHEX-MATE subsample used for this analysis is shown in Fig.~\ref{fig:gallery}, sorted with increasing values of the centroid shift $w$. 

\label{app:gallery}

\end{appendix}
\end{document}